\documentclass[aps, prl, twocolumn, super0scriptaddress, floatfix, longbibliography, nofootinbib, bibnotes]{revtex4-2}

\pdfoutput=1
\usepackage{amsmath}
\usepackage{amssymb}
\usepackage{graphicx}
\usepackage{booktabs}
\usepackage{multirow}
\usepackage{placeins}
\usepackage{capt-of}
\usepackage{array}
\newcolumntype{L}[1]{>{\raggedright\arraybackslash}p{#1}}

\usepackage[colorlinks=true,breaklinks=true]{hyperref}
\usepackage[dvipsnames]{xcolor}
\definecolor{darkred}{rgb}{0.5,0,0}
\definecolor{darkblue}{rgb}{0,0,0.5}
\definecolor{firebrick}{rgb}{0.75,0.125,0.125}
\definecolor{darkgreen}{rgb}{0,0.5,0}
\hypersetup{urlcolor=darkblue,
	    citecolor=darkgreen,
	    linkcolor=firebrick}

\usepackage{orcidlink}

\graphicspath{{figures/}}

\newcommand{\ie}{{\it i.e.}}

\newcommand{\eg}{{\it e.g.}}

\newcommand{\eq}{Eq.}

\newcommand{\fig}{Fig.}

\providecommand{\Ref}{}
\AtBeginDocument{\renewcommand{\Ref}{Ref.}}
\newcommand{\Refs}{Refs.}

\newcommand{\equ}[1]{\eq~(\ref{equ:#1})}
\newcommand{\figu}[1]{\fig~\ref{fig:#1}}

\newcommand{\fE}[1]{f_{#1,\oplus}}
\newcommand{\fS}[1]{f_{#1,{\rm S}}}

\begin{document}

\title{Measuring the flavor composition of high-energy neutrinos from the Milky Way}

\author{Mauricio Bustamante\,\orcidlink{0000-0001-6923-0865}}
\email{mbustamante@nbi.ku.dk}
\affiliation{Niels Bohr International Academy, Niels Bohr Institute,\\University of Copenhagen, 2100 Copenhagen, Denmark}

\date{\today}

\begin{abstract}
The origin of the Milky Way's high-energy neutrinos is debated.
Their flavor composition reflects how they are made and could help settle it, but remains to be measured.
We show that IceCube and KM3NeT, which see the inner Galaxy in complementary ways from opposite hemispheres, could start resolving its flavor composition in about a decade.
With upcoming telescopes, they would infer the $\nu_e$ fraction at the sources to within 27\%.
A first comparison of Galactic and extragalactic flavor compositions could reveal strong magnetic fields in extragalactic sources or neutrino decay.
\end{abstract}

\maketitle

\smallskip
\textit{Introduction.}---How are the high-energy neutrinos from the Milky Way made?
After a hint from ANTARES~\cite{ANTARES:2022izu}, IceCube found evidence for TeV-scale neutrinos from the Galactic plane~\cite{IceCube:2023ame, Bustamante:2023iyn} and has since observed them at more than $5\sigma$~\cite{IceCube:2026plr}.
IceCube has also measured how their flux and spectral index vary along the plane, in segments of Galactic longitude~\cite{IceCube:2026hqk}.
Their origin is still debated: they may come from cosmic rays colliding with interstellar gas~\cite{Stecker:1978ah, Gaggero:2015xza, TibetASgamma:2021tpz, Schwefer:2022zly, LHAASO:2023gne, DeLaTorreLuque:2025zsv} or from unresolved sources~\cite{IceCube:2023ame, Fang:2023azx, Roberts:2026cjf}.

Their flavor composition, \ie, the fractions of $\nu_e$, $\nu_\mu$, and $\nu_\tau$ in their flux, has not been measured~\cite{IceCube:2026plr}, though it is set by how the neutrinos are made~\cite{Learned:1994wg, Athar:2000yw, Beacom:2003nh, Kashti:2005qa, Lipari:2007su, Bustamante:2015waa, Bustamante:2019sdb, Song:2020nfh}.
Cosmic-ray collisions with gas make charged pions, whose decays, via muons, yield $\nu_e$ and $\nu_\mu$ in ratio 1:2 and no $\nu_\tau$.
Unresolved sources would change this ratio if they made neutrinos differently, \eg, in magnetic fields strong enough to cool the muons.
Thus, measuring the flavor composition tests whether the Milky Way's neutrinos are made as expected.

Once measured, the plane's flavor composition would also serve as a reference for that of the extragalactic flux, \ie, the high-energy neutrinos from beyond the Milky Way, whose composition IceCube has measured only imprecisely~\cite{IceCube:2015rro, IceCube:2020fpi, IceCube:2024nhk, IceCube:2025ole}.
This composition probes high-energy astrophysics~\cite{Rachen:1998fd, Athar:2000yw, Barenboim:2003jm, Beacom:2003nh, Kashti:2005qa, Mena:2006eq, Kachelriess:2006ksy, Lipari:2007su, Esmaili:2009dz, Choubey:2009jq, Hummer:2010ai, Winter:2013cla, Palladino:2015zua, Bustamante:2015waa, Biehl:2016psj, Bustamante:2019sdb, Ackermann:2019ows, Bustamante:2020bxp, Song:2020nfh, Bhattacharya:2023mmp, Telalovic:2023tcb, Dev:2023znd} and fundamental physics~\cite{Crocker:2001zs, Beacom:2002vi, Barenboim:2003jm, Beacom:2003nh, Beacom:2003eu, Beacom:2003zg, Serpico:2005bs, Mena:2006eq, Lipari:2007su, Pakvasa:2007dc, Esmaili:2009dz, Choubey:2009jq, Esmaili:2009fk, Bhattacharya:2009tx, Bhattacharya:2010xj, Bustamante:2010nq, Mehta:2011qb, Baerwald:2012kc, Fu:2012zr, Pakvasa:2012db, Chatterjee:2013tza, Xu:2014via, Aeikens:2014yga, Arguelles:2015dca, Bustamante:2015waa, Pagliaroli:2015rca, Shoemaker:2015qul, deSalas:2016svi, Gonzalez-Garcia:2016gpq, Bustamante:2016ciw, Rasmussen:2017ert, Dey:2017ede, Bustamante:2018mzu, Farzan:2018pnk, Ahlers:2018yom, Brdar:2018tce, Palladino:2019pid, Ackermann:2019cxh, Arguelles:2019rbn, Ahlers:2020miq, Karmakar:2020yzn, Fiorillo:2020gsb, Song:2020nfh, IceCube:2021tdn, Arguelles:2022tki, MammenAbraham:2022xoc, Telalovic:2023tcb, Telalovic:2025xor}, including neutrino mixing at high energies~\cite{Bustamante:2026aur, Bustamante:2026zst, Bustamante:2026fxa}.

Galactic and extragalactic neutrinos have so far been compared through their intensities and spectra~\cite{Fang:2023azx, IceCube:2026plr}.
Flavor adds a comparison of a different kind, since the two compositions differ only if the neutrinos are made, or travel, differently~\cite{Telalovic:2023tcb, Telalovic:2025xor}.
The extragalactic ones would differ if muons lost energy in strong magnetic fields at their sources~\cite{Kashti:2005qa, Hummer:2010ai, Winter:2013cla, Bustamante:2020bxp}, or if new physics, \eg, neutrino decay~\cite{Beacom:2002vi, Baerwald:2012kc, Bustamante:2016ciw}, changed their flavor over cosmological but not Galactic distances.
Earlier work proposed testing new physics with Galactic neutrinos alone, assuming that pion decay makes them~\cite{MacDonald:2025jbm}; measuring their composition would test that assumption.

\begin{figure*}[t]
  \centering
  \includegraphics[width=\textwidth]{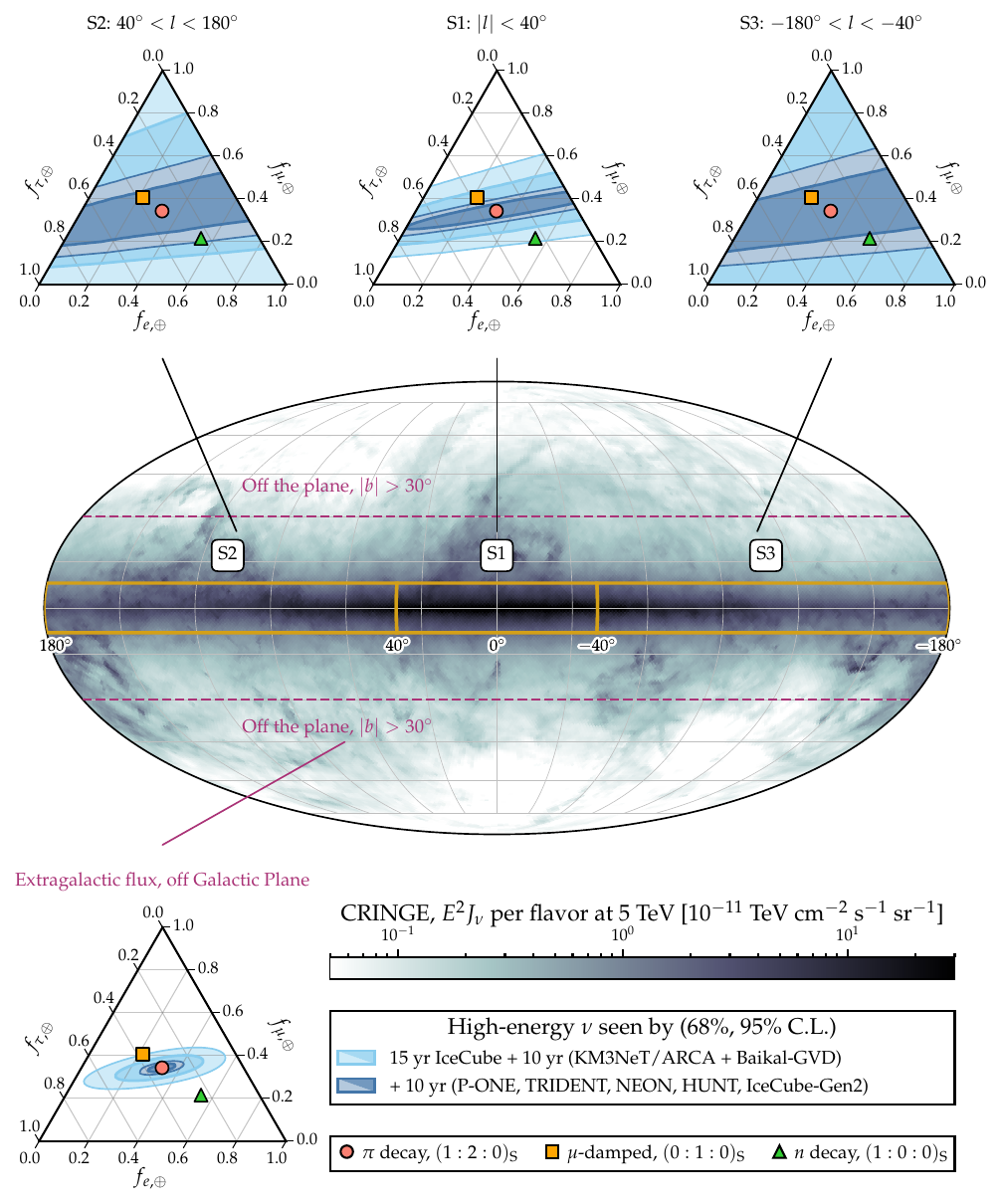}
  \caption{\label{fig:headline}\textbf{\textit{Flavor composition of the high-energy neutrinos along the Galactic plane and from outside it.}}
  Forecast measurements of the composition at Earth in IceCube's three segments of the plane and of the extragalactic flux, all fitted together with every systematic uncertainty.
  The segments span ranges of Galactic longitude, $l$, at Galactic latitudes $|b| < 8^\circ$.
  The sky map shows the CRINGE flux~\cite{Schwefer:2022zly} for illustration only.
  Our forecast takes each segment as uniformly bright, at IceCube's measured flux and spectral index~\cite{IceCube:2026hqk}.
  In every segment and in the extragalactic flux, the sources make their neutrinos by pion decay.
  See the text for details.
  The inner segment, the brightest, is measured best.}
\end{figure*}

We forecast how precisely neutrino telescopes now running or being built (IceCube, KM3NeT, Baikal-GVD, P-ONE) and planned (TRIDENT, NEON, HUNT, IceCube-Gen2) can measure the flavor composition along the Galactic plane (\figu{headline}).
IceCube sees the inner Galaxy mostly in showers from $\nu_e$ and $\nu_\tau$, while KM3NeT sees most of the plane in tracks from $\nu_\mu$.
\textbf{With IceCube's 15 years of data and 10 years of KM3NeT at full size, the two could tell the inner segment's flavor composition at about $2\sigma$ from the $\nu_e$-rich one expected from neutron decay.}
With 10 years of the other six, all eight would measure the $\nu_e$ fraction at the inner segment's sources to within 27\%, telling pion decay apart from both neutron decay and sources with strong magnetic fields.
\textbf{The measurement would test how these neutrinos are made and open a new way to compare Galactic and extragalactic neutrinos.}

\smallskip
\textit{Flavor at the sources and at Earth.}---Cosmic rays make charged pions in collisions with gas ($pp$)~\cite{Stecker:1978ah, Kelner:2006tc} or with photons ($p\gamma$)~\cite{Stecker:1978ah, Kelner:2008ke, Hummer:2010vx}.
Diffuse production in the plane's interstellar gas is $pp$, while unresolved and extragalactic sources may use either process.
Pions decay as $\pi^+ \to \mu^+ + \nu_\mu$, then $\mu^+ \to e^+ + \nu_e + \bar\nu_\mu$, or through the charge-conjugate chain.
This yields flavor fractions at production $\fS{\alpha}$, \ie, ratios of the flux of $\nu_\alpha + \bar\nu_\alpha$ to the total, of $(\fS{e}, \fS{\mu}, \fS{\tau}) = (1/3, 2/3, 0)$~\cite{Learned:1994wg, Lipari:2007su}.
We call $\fS{\alpha}$ the fractions at the sources (even if diffuse).
We assume equal $\nu_\alpha$ and $\bar\nu_\alpha$ fluxes, as $pp$ collisions give~\cite{Kelner:2006tc}.

Muons lose energy to synchrotron radiation before decaying  for neutrinos with energy above $E_c \simeq 2 \times 10^9~{\rm GeV}\,(B/{\rm G})^{-1}$, where $B$ is the magnetic field in the production region (End Matter)~\cite{Kashti:2005qa, Hummer:2010ai, Bustamante:2020bxp}, \ie, above about $10^{15}$~GeV in the few-$\mu$G interstellar fields of the Milky Way~\cite{Jansson:2012pc}, far above the observed energies.
In fields above about 20~kG, $E_c$ is below 100~TeV, among the observed energies.
When muons cool, only the $\nu_\mu$ from the pion decays keep high energies, which gives the muon-damped composition, $(0, 1, 0)_{\rm S}$~\cite{Rachen:1998fd, Kashti:2005qa, Kachelriess:2006ksy, Kachelriess:2007tr, Hummer:2010ai}.
Neutrons that decay in flight, $n \to p + e^- + \bar\nu_e$, proposed as a Galactic source of TeV $\bar\nu_e$~\cite{Anchordoqui:2003vc}, give $(1, 0, 0)_{\rm S}$.
Unlike pion decay and muon damping, neutron decay makes only antineutrinos, which the telescopes detect less often than neutrinos.
Every result we quote against neutron decay holds for a flux of $\bar\nu_e$ only (the significance is lower than with equal fluxes of $\nu_e$ and $\bar\nu_e$; Supplemental Material).

Pion decay, muon damping, and neutron decay are benchmarks that a flux from sources of different kinds would mix.
Within one source, the composition moves from pion decay below $E_c$ to muon damping above it, and changes again where pions, too, cool~\cite{Kashti:2005qa, Lipari:2007su, Hummer:2010ai}.
We take the flavor composition as constant, \ie, averaged over the detected energies, as is standard, since energy dependence is challenging to detect~\cite{Liu:2023flr}.

Neutrino oscillations change the flavor composition en route to Earth~\cite{ParticleDataGroup:2024cfk}.
Over in-Galaxy kiloparsec and cosmological distances, oscillations average out~\cite{Learned:1994wg}, since even at 10~PeV the longest oscillation length is only about 0.01~pc.
The $\nu_\beta+\bar\nu_\beta$ fraction at Earth is then~\cite{Bustamante:2015waa, Song:2020nfh, Liu:2023flr} $\fE{\beta} = \sum_{\alpha=e,\mu,\tau} P_{\alpha\beta}\, \fS{\alpha}$, where $P_{\alpha\beta} = \sum_{i=1}^{3} |U_{\alpha i}|^2 |U_{\beta i}|^2$ is the probability that a $\nu_\alpha$ reaches Earth as a $\nu_\beta$, and $U$ is the lepton mixing matrix.
We evaluate it at the best-fit values of the mixing parameters on which $U$ depends, from the NuFIT~6.1 global fit assuming normal mass ordering~\cite{Esteban:2024eli, NuFIT61}, with their reduced projected uncertainties from \Ref~\cite{Bustamante:2026fxa}.
At Earth, pion decay gives nearly equal fractions, $(0.33, 0.34, 0.33)_\oplus$ (\figu{headline}).
Production of $\nu_\tau$ is negligible, so we describe the sources by $(\fS{e}, 1-\fS{e}, 0)$.

\smallskip
\textit{Telescopes.}---Neutrino telescopes collect the Cherenkov light of the charged particles made when high-energy neutrinos scatter, mainly on nucleons, in water or ice~\cite{Cooper-Sarkar:2011jtt, IceCube:2016zyt, Katori:2021nwq, Spiering:2026hqf}.
The light profile tells flavors apart~\cite{Beacom:2003nh}: muons from $\nu_\mu$ charged-current interactions leave elongated tracks that point back to the neutrino direction, whereas charged-current interactions of $\nu_e$ and $\nu_\tau$ and neutral-current ones of any flavor make showers, compact profiles with poorer pointing.
Rarely, a $\nu_\tau$ makes a double cascade, two showers from its interaction followed by a tau decay~\cite{Learned:1994wg}.
We consider all three event types.
Since tracks from atmospheric muons arriving from above outnumber neutrino tracks, we keep only tracks from below or near each telescope's horizon.
The inner Galaxy lies mostly in the southern sky, which IceCube, at the South Pole, sees almost only in showers.
A telescope in the Northern Hemisphere, such as KM3NeT, complements it with up-going tracks. 
More telescopes worldwide add events of both kinds~\cite{Schumacher:2021hhm, Schumacher:2025qca}.

We consider the telescopes now running or being built, IceCube~\cite{IceCube:2016zyt}, KM3NeT~\cite{KM3Net:2016zxf}, Baikal-GVD~\cite{Baikal-GVD:2025rhg}, and P-ONE~\cite{P-ONE:2020ljt}, and the planned TRIDENT~\cite{TRIDENT:2022hql}, NEON~\cite{Zhang:2024slv}, HUNT~\cite{Huang:2023mzt}, and IceCube-Gen2~\cite{IceCube-Gen2:2020qha}.
The Supplemental Material lists their sites and sizes.
Our main results combine either IceCube, KM3NeT, and Baikal-GVD, the existing telescopes, or all eight.
From IceCube, we use the Galactic-plane cascade sample with which it observed the plane~\cite{IceCube:2023ame, IceCube:GalacticPlaneData} and its through-going tracks, IceTracks-DR2~\cite{Abbasi:2026ehs, IceCube:DR2Data}; from every other telescope, tracks, showers, and double cascades.
IceCube contributes 15 years of data and every other telescope 10 years at full size.
The other telescopes' responses, modeled as closely as public information allows, start from IceTracks-DR2 for tracks and from IceCube's medium-energy starting events~\cite{IceCube:MESEflavorData} for showers and double cascades, carried to each site and rescaled to each size, then completed with, and checked against, what each collaboration has published (Supplemental Material).
Our model of IceCube's cascade sample reproduces its measurement along the plane~\cite{IceCube:2026hqk}: in the inner Galaxy, at $|l| < 30^\circ$, it gives a 68\%~C.L.\ interval of $\pm32\%$ on the flux normalization at 5~TeV, against IceCube's $-29\%/+32\%$.

\smallskip
\textit{Analysis.}---We forecast median sensitivities on Asimov data~\cite{Cowan:2010js}, \ie, the expected event counts for assumed true fluxes.
In each of IceCube's three segments of the Galactic plane (\figu{headline}), at Galactic latitudes $|b| < 8^\circ$, the true flux is a power law in energy, $\Phi \propto E^{-\gamma}$, with the normalization and spectral index $\gamma$ that IceCube fitted~\cite{IceCube:2026hqk}.
The extragalactic flux is isotropic, with the broken power law of IceCube's combined fit of tracks and showers~\cite{IceCube:2025ewu}.
Both have pion decay at the sources; atmospheric neutrinos~\cite{Fedynitch:2015zma, Gaisser:2011klf, Riehn:2019jet} and muons are the backgrounds.

We fit the three segments together with the sky away from the plane, at $|b| > 30^\circ$, where the extragalactic flux dominates and is measured.
The fit minimizes a binned Poisson $\chi^2$, summed over every telescope and event type, with confidence regions from Wilks' theorem~\cite{Wilks:1938dza} (Supplemental Material).
Each segment's normalization, spectral index, and flavor composition are free, as is the separate extragalactic composition.
Up to 55 nuisance parameters, most with priors, describe the systematic uncertainties in the extragalactic spectrum, the atmospheric fluxes, and each telescope's efficiency, energy scale, pointing, and event misidentification.

\smallskip
\textit{Flavor along the plane.}---Figure~\ref{fig:headline} shows the compositions at Earth that the telescopes would measure in the three segments and in the extragalactic flux.
Each confidence region is a long, narrow band along which the $\nu_e$ and $\nu_\tau$ fractions vary at a nearly fixed $\nu_\mu$ fraction, because tracks fix the $\nu_\mu$ content while showers barely tell $\nu_e$ from $\nu_\tau$.
More tracks make the band narrower.
The inner segment is measured best, since it is about four to five times brighter per steradian than the other two segments~\cite{IceCube:2026hqk} and lies where the northern telescopes see its tracks.
The extragalactic flux is measured better still, since the sky away from the plane holds about five times more astrophysical neutrinos than the inner segment.

\begin{figure}[t]
  \centering
  \includegraphics[width=\columnwidth]{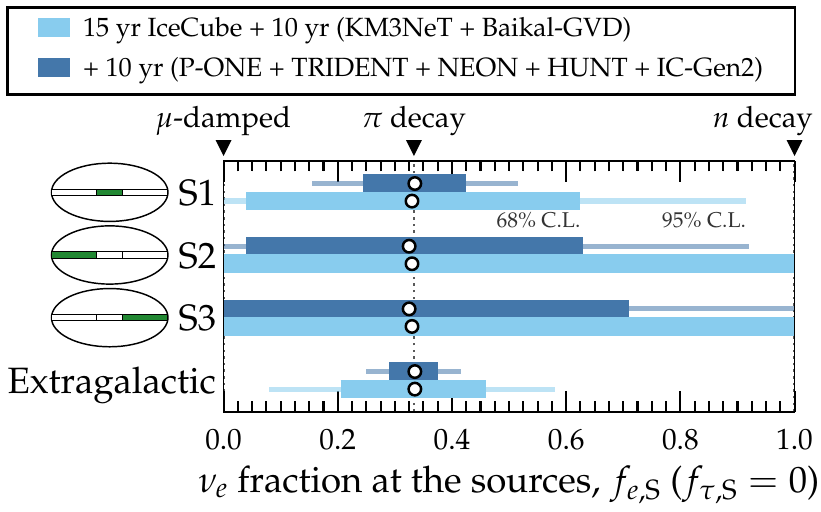}
  \caption{\label{fig:fes}\textbf{\textit{$\nu_e$ fraction at the sources of the Galactic-plane segments and of the extragalactic flux.}}
  Inferred from the compositions of \figu{headline} for sources without $\nu_\tau$, with the mixing parameters free within their uncertainties projected for 2040~\cite{Bustamante:2026fxa}.
  Pion decay, $\fS{e} = 1/3$, is the truth in all segments and in the extragalactic flux.
  See the text for details.
  Above 95\%~C.L., pion decay is told apart in the inner segment from neutron-decay sources by the existing telescopes and also from muon-damped sources by all eight.}
\end{figure}

Figure~\ref{fig:fes} shows the $\nu_e$ fraction at the sources, $\fS{e}$, which we infer from these measurements for sources without $\nu_\tau$.
For each value of $\fS{e}$, we compute the composition at Earth and take its $\Delta\chi^2$ in the measurement of \figu{headline}, minimized over the mixing parameters.
The resulting profile of $\Delta\chi^2$ in $\fS{e}$ alone gives its C.L.\ intervals, in a frequentist version of the method introduced in \Ref~\cite{Bustamante:2019sdb} and applied in \Refs~\cite{Song:2020nfh, IceCube:2025ole}.

With pion decay at the sources, all eight telescopes would measure the inner segment's $\fS{e}$ to within 27\% of its true value, 1/3, at 68\%~C.L.
At this precision, pion decay is told apart from both muon damping and neutron decay.
The existing telescopes would measure the inner segment's $\fS{e}$ to within 88\%, enough to disfavor neutron decay but not muon damping.
The extragalactic $\fS{e}$, measured from more events, is about twice as precise, to within 13\% with all eight telescopes and 38\% with the existing ones.
IceCube and KM3NeT alone measure $\fS{e}$ nearly as precisely as with Baikal-GVD (to within about 90\% in the inner segment either way).
With all eight telescopes, the precision in the inner segment barely changes if the true fluxes are those predicted by CRINGE~\cite{Schwefer:2022zly} (30\% instead of 27\%).
IceCube also measured the plane in six segments~\cite{IceCube:2026hqk}.
If we fit them instead of the three, the inner one is measured as precisely (Supplemental Material).

The Galactic plane's composition would approach muon damping only if most of its neutrinos came from sources whose magnetic fields cool the muons.
Microquasar jets may hold such fields near the compact object, but the neutrinos made there are hard to detect~\cite{Reynoso:2008gs}.
Those made farther out along the jet~\cite{Reynoso:2008gs}, or in interstellar gas by the cosmic rays that microquasars accelerate~\cite{Zhang:2026eju}, keep the composition of pion decay.
With all eight telescopes, though not with the existing ones, an inner segment at muon damping would be told from pion decay at about $3\sigma$ (Supplemental Material).

\begin{figure}[t]
  \centering
  \includegraphics[width=\columnwidth]{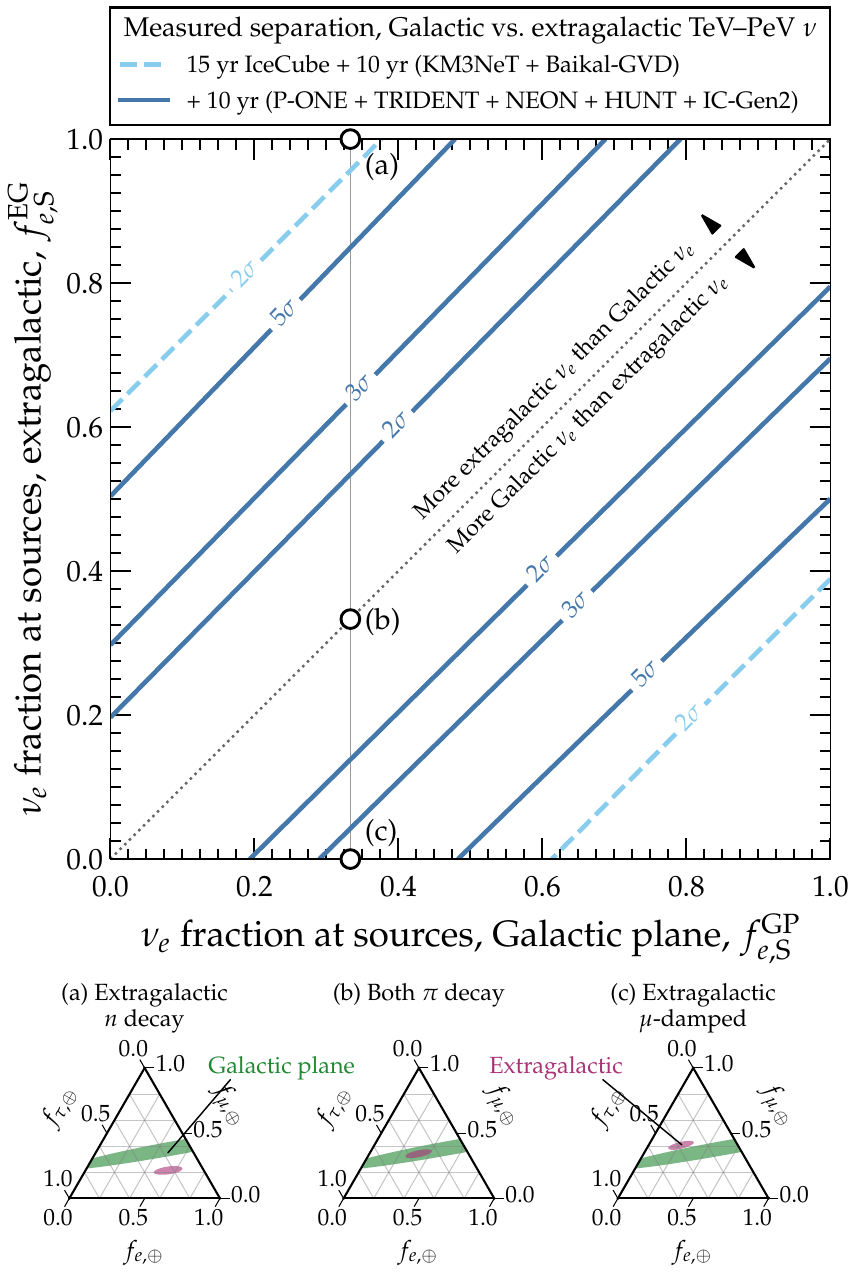}
  \caption{\label{fig:separation}\textbf{\textit{Telling the $\nu_e$ fraction at the sources of the Galactic plane from that of the extragalactic flux.}}
  Significance with which the data reject a single $\nu_e$ fraction at the sources common to the Galactic plane and to the extragalactic flux, for each pair of true fractions.
  The triangles show the compositions at Earth that all eight telescopes would measure at the marked pairs (95\%~C.L.).
  See the text for details.}
\end{figure}

\smallskip
\textit{Galactic vs.~extragalactic.}---Figure~\ref{fig:separation} shows how different the $\nu_e$ fractions at the sources of the Galactic plane and of the extragalactic flux must be for the telescopes to tell them apart.
For each pair of true values, we compute how strongly the data reject one composition common to both fluxes, with all spectra and nuisance parameters free and the three segments sharing one composition.
The significance with which the fractions are told apart grows in proportion to their difference, whatever their values.
It rises by about $1\sigma$ for every 0.1 of difference with all eight telescopes, but only for every 0.3 with the existing ones.
The triangles show why: changing the extragalactic $\fS{e}$ moves its composition at Earth across the plane's narrow band, farther for neutron decay than for muon damping.

The comparison may reveal muon cooling in extragalactic sources, which moves their composition, but not the plane's, towards muon damping above $E_c$~\cite{Kashti:2005qa, Kachelriess:2007tr, Hummer:2010ai, Bustamante:2020bxp, Liu:2023flr} (End Matter).
If the sources emit a broken power law, favored by IceCube~\cite{IceCube:2025ewu}, all eight telescopes detect cooling at $3\sigma$ for source populations with average magnetic fields above about 11~kG and at $5\sigma$ above 22~kG;  existing telescopes reach at most $2.4\sigma$ below 300~kG.
Several source classes can host fields this strong~\cite{Bustamante:2020bxp}: low-luminosity~\cite{Ghisellini:2006ng} and high-luminosity~\cite{Piran:2005qu} gamma-ray bursts and neutron-star and magnetar winds~\cite{Arons:2002yj}.

\smallskip
\textit{Testing new physics.}---Comparing the plane with the extragalactic flux can also reveal new physics that acts over cosmological distances, such as the decay of neutrino mass eigenstates $\nu_2$ and $\nu_3$~\cite{Beacom:2002vi, Baerwald:2012kc, Pagliaroli:2015rca, Bustamante:2016ciw} (End Matter).
In this case, decaying extragalactic neutrinos reach Earth with the composition of the remaining stable $\nu_1$, while the plane's neutrinos, after only a few kpc, barely decay for lifetime-to-mass ratios, $\tau/m$, above about 1~s~eV$^{-1}$.
At shorter lifetimes, the plane's own neutrinos decay, which IceCube and KM3NeT could probe with the plane alone~\cite{MacDonald:2025jbm}.
Present extragalactic data disfavor $\tau/m \lesssim 220$~s~eV$^{-1}$ at 95\%~C.L.~\cite{Valera:2024buc}, so the comparison matters only at longer lifetimes.
At $2\sigma$, it would reach about $3 \times 10^3$~s~eV$^{-1}$ with the existing telescopes, close to the 2035 projection for the extragalactic flux alone~\cite{Valera:2024buc}.
All eight would probe lifetimes up to about $1.5 \times 10^4$~s~eV$^{-1}$, seventy times longer than today.

\smallskip
\textit{Outlook.}---The Milky Way's high-energy neutrinos have been seen, but their origin is still debated.
We have forecast, for the first time, how well neutrino telescopes can measure their flavor composition, which tests how they are made and so helps probe their origin.

The plane's neutrinos are expected to come from the decays of charged pions, which make $\nu_e$ and $\nu_\mu$ in the ratio 1:2.
\textbf{IceCube and KM3NeT, both running, see the inner Galaxy mainly in showers and in tracks, respectively, and together could tell neutrino production from pion decay apart, at about $2\sigma$ after 10 years of KM3NeT at full size, from production via neutron decay, which makes only $\bar\nu_e$.
Aided by upcoming telescopes, they would measure the inner Galaxy's $\nu_e$ fraction at the sources to within 27\% of its value, enough to tell pion decays apart, at over 95\%~C.L., from both decaying neutrons and sources with strong magnetic fields that make only $\nu_\mu$.}
Comparing the plane's composition with that of extragalactic neutrinos would reveal muon cooling in their sources for fields above about $10$~kG.
It would also reveal the decay of $\nu_2$ and $\nu_3$ over cosmological distances for lifetime-to-mass ratios up to about $1.5 \times 10^4$~s~eV$^{-1}$, far beyond present limits.

Techniques under study to improve flavor separation, using the delayed light after showers~\cite{Li:2016kra, Steuer:2017tca, Farrag:2023jut, IceCube:2025paj} or starting tracks~\cite{Wen:2026ngx}, would sharpen the measurement beyond our forecasts.
Longer exposure could establish whether the composition varies along the plane~\cite{Telalovic:2023tcb} and how it changes with energy~\cite{Liu:2023flr}.
Measuring the plane's composition would make it a flavor reference, measured rather than assumed, against which the extragalactic composition would reveal how distant sources make neutrinos and what happens to them over cosmological distances.

\newpage
\smallskip
\textit{Acknowledgments.}---MB is supported by {\sc Villum Fonden} under project no.~29388.
This work used the Tycho supercomputer hosted at the SCIENCE High Performance Computing Center at the University of Copenhagen.
During the preparation of this work, the author used Claude for research assistance in code development, numerical auditing, figure preparation, and editorial revision.
All scientific judgments, validation decisions, and responsibility for the manuscript remain with the author.

%


\clearpage
\appendix
\section*{End Matter}

\section{Cooling in the extragalactic sources}
\label{app:cooling}

A muon in a magnetic field $B$ loses energy by synchrotron radiation faster than it decays above the energy
\begin{equation}
  \label{equ:ecmu}
  E_{c,\mu} = \left(\frac{6\pi\, m_\mu^5}{\sigma_{\rm T}\, m_e^2\, \tau_\mu\, B^2}\right)^{1/2}
  \simeq 6\times10^{9}~{\rm GeV}\left(\frac{B}{\rm G}\right)^{-1} ,
\end{equation}
where $m_e$ is the electron mass, $\sigma_{\rm T}$ the Thomson cross section, and $m_\mu$ and $\tau_\mu$ the muon mass and lifetime~\cite{Kashti:2005qa, Hummer:2010ai, Bustamante:2020bxp}.
Neutrinos from muon decay carry about a third of the muon's energy, so they are suppressed above $E_c \approx E_{c,\mu}/3$.
Pions and kaons, which decay faster, cool at higher energies~\cite{Kachelriess:2007tr, Hummer:2010ai} (Supplemental Material).

The true extragalactic flux is a broken or simple power law, cooled in sources with field $B$ and fitted to IceCube's measured spectrum~\cite{IceCube:2025ewu}.
The alternative hypothesis has this form, with $B$ and the spectrum free (Supplemental Material).
The null hypothesis has no cooling, one composition for the plane and the extragalactic flux, free and constant in energy, and one more break in its spectrum than at the sources, to mimic the bend from cooling.

Figure~\ref{fig:cooling} shows how strongly the data reject the null hypothesis, and the compositions at three fields.
With a broken power law, the significance peaks when about half of the extragalactic events lie above $E_c$, so that the compositions below and above $E_c$ are both measured from many events~\cite{Liu:2023flr}.
With a simple power law, more events are detected and the cooled spectrum departs more from one with a single break, so the significance is higher.

\vspace{-8pt}
\section{Neutrino decay}
\label{app:decay}

If $\nu_2$ and $\nu_3$ decay into undetected particles, \ie, invisibly, with equal lifetime-to-mass ratios $\tau/m$, the probability that each reaches Earth with energy $E$ is
\begin{equation}
  \label{equ:survival}
  S(E) = \exp\!\left(-\frac{m}{\tau}\,\frac{L_{\rm eff}}{E}\right) ,
\end{equation}

\addtocounter{figure}{1}
\begin{figure*}[t]
  \centering
  \includegraphics[width=\textwidth]{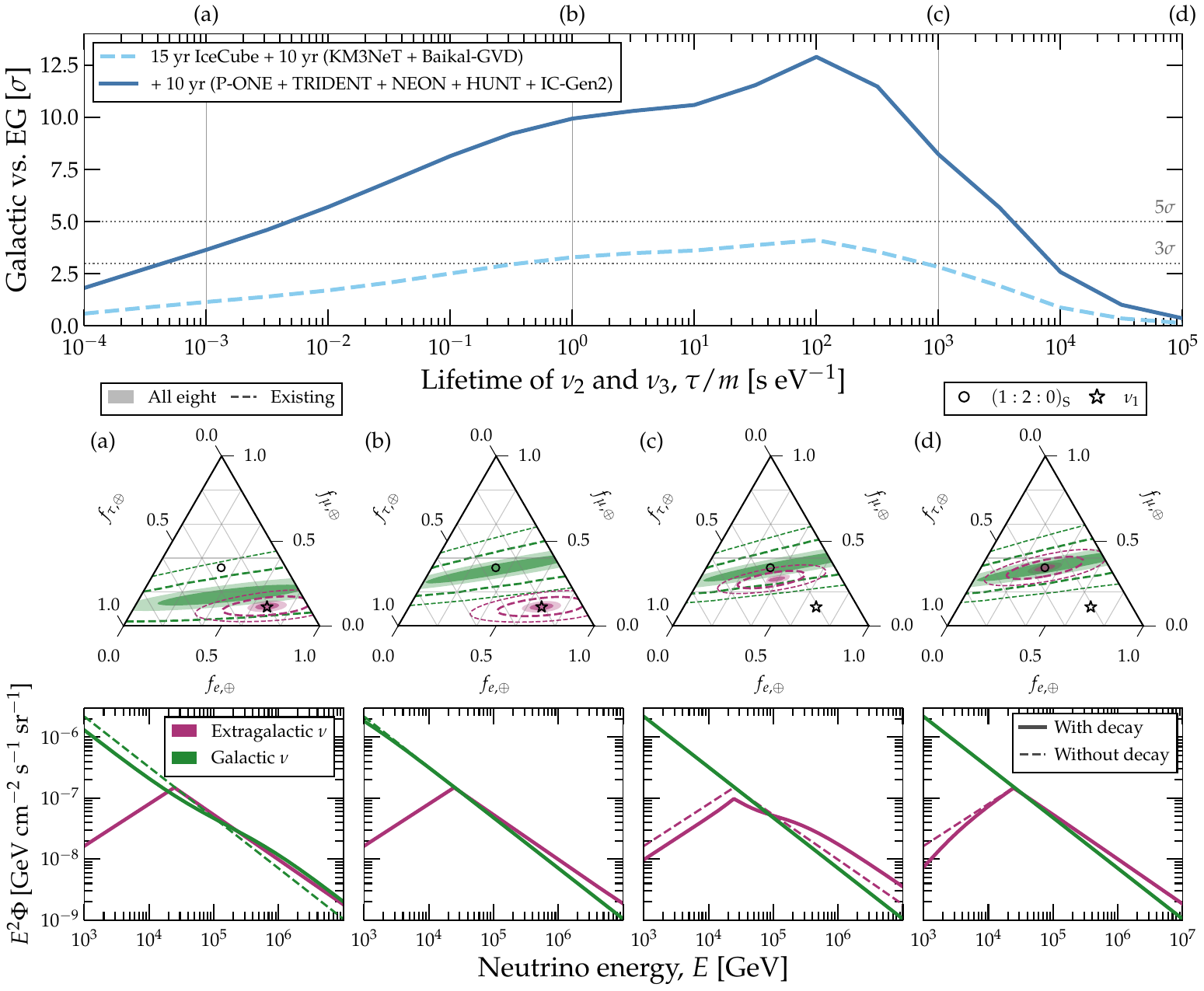}
  \caption{\label{fig:decay}\textbf{\textit{Invisible decay of $\nu_2$ and $\nu_3$: telling the Galactic plane from the extragalactic flux.}}
  \textit{Top:} significance of rejecting one flavor composition, constant in energy, shared by the Galactic plane and the extragalactic (EG) flux.
  \textit{Middle:} flavor compositions at Earth measured at the four marked lifetimes, with the three segments sharing one (68\% and 95\%~C.L., dark and light).
  \textit{Bottom:} all-flavor intensities of the inner segment and the extragalactic flux, those with and without decay normalized to each other between 10~TeV and 1~PeV.
  See the text for details.
  With all eight telescopes, the two compositions differ at more than $3\sigma$ over more than seven decades of lifetime.}
\end{figure*}
\addtocounter{figure}{-2}

\onecolumngrid
\vspace*{12pt}
\noindent\begin{minipage}{\textwidth}
  \centering
  \includegraphics[width=\textwidth]{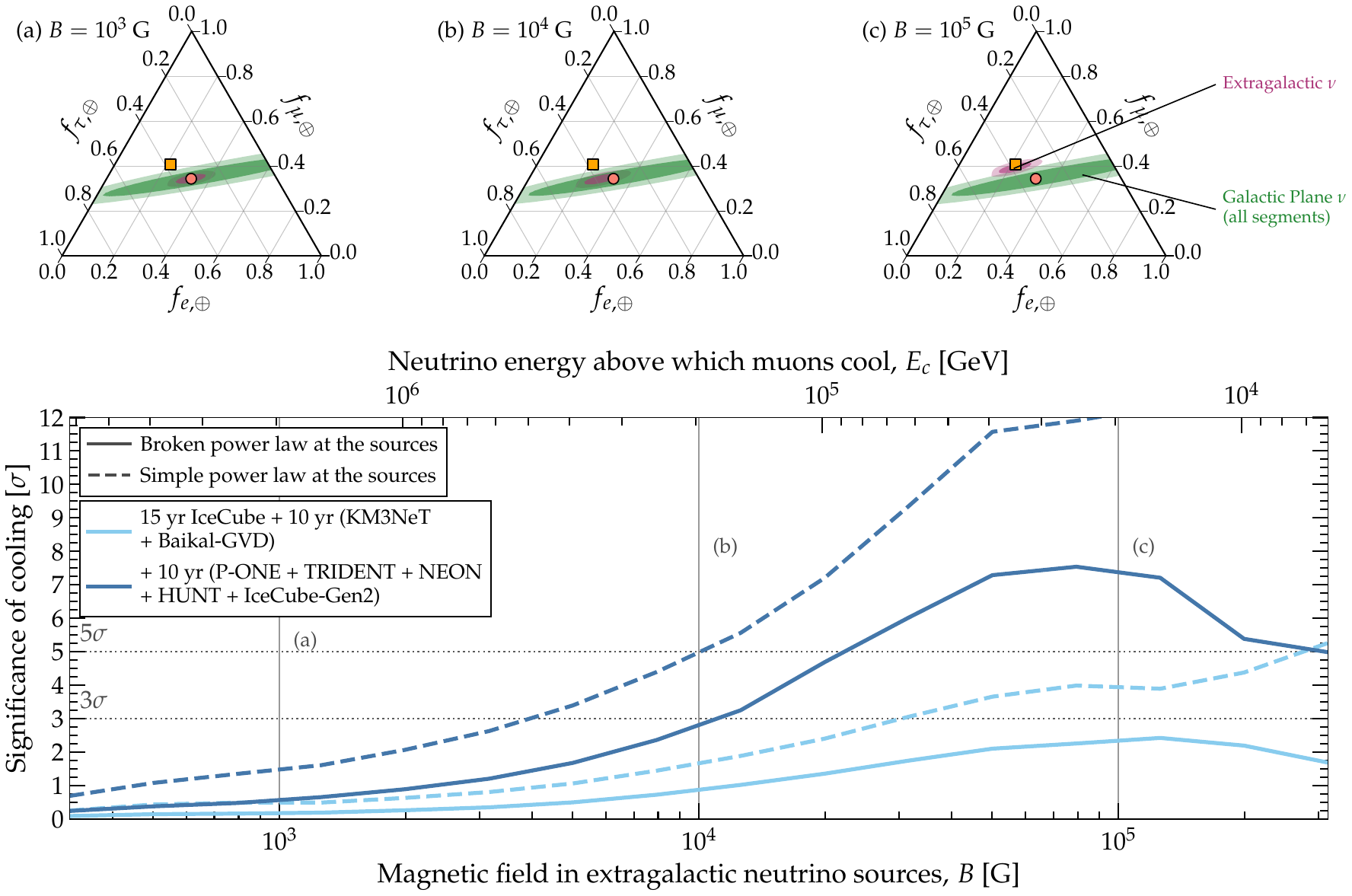}
  \captionof{figure}{\label{fig:cooling}\textbf{\textit{Significance of muon cooling in the extragalactic sources, against their magnetic field.}}
  \textit{Top:} compositions at Earth measured by all eight telescopes at the three marked fields, for a broken power law at the sources (68\% and 95\%~C.L., dark and light), and those of pion decay (circle) and muon damping (square).
  \textit{Bottom:} significance of cooling, with $B$ free, against no cooling with one composition for the plane and the extragalactic flux, constant in energy.
  See the text for details.
  With all eight telescopes, cooling in fields above about $10^4$~G is detected at more than $3\sigma$ for either spectrum.}
\end{minipage}
\addtocounter{figure}{1}
\twocolumngrid
\noindent where $L_{\rm eff}$ is the distance traveled in the Milky Way, or $\int_0^z dz'/[(1+z')^2 H(z')]$ from a source at redshift $z$, with $H$ the Hubble rate.
Since oscillations average out over these distances, the flux of each flavor at Earth sums the surviving mass eigenstates, weighted by their content of that flavor~\cite{Beacom:2002vi}.
The density of extragalactic sources follows the star-formation rate up to $z = 4$~\cite{Hopkins:2006bw, Yuksel:2008cu, Kistler:2009mv}; the plane's sources fill a thin disc (Supplemental Material).
Under the null hypothesis, the plane and the extragalactic flux share one composition, free and constant in energy, while the extragalactic spectrum is a free broken power law.

Figure~\ref{fig:decay} shows how strongly the data reject the null hypothesis, and the flavor compositions and fluxes at four lifetimes.
The significance is largest where the extragalactic neutrinos have decayed at the detected energies, leaving only $\nu_1$ at Earth, but the plane's have not.
It falls at shorter lifetimes, where the plane's neutrinos decay too, and at longer ones, where the extragalactic neutrinos decay only at low energies.
If only $\nu_3$ decays, the extragalactic composition changes less, so that all eight telescopes reach at most $3.5\sigma$.
Complete visible decay, into detected $\nu_1$~\cite{Abdullahi:2020rge}, is told from no decay about as well as complete invisible decay (Supplemental Material).


\clearpage
\begin{center}
\textbf{\large Supplemental Material}
\end{center}


\setcounter{section}{0}
\setcounter{secnumdepth}{3}
\renewcommand{\thesection}{\Alph{section}}
\counterwithin*{figure}{section}
\counterwithin*{table}{section}
\counterwithin*{equation}{section}
\renewcommand{\thefigure}{\thesection.\arabic{figure}}\renewcommand{\theHfigure}{\thesection.\arabic{figure}}
\renewcommand{\thetable}{\thesection.\arabic{table}}\renewcommand{\theHtable}{\thesection.\arabic{table}}
\renewcommand{\theequation}{\thesection\arabic{equation}}\renewcommand{\theHequation}{\thesection\arabic{equation}}

\section{The telescopes and their responses}
\label{sm:telescopes}

A telescope's response gives, for each flavor, energy, and direction of a neutrino, the expected numbers of tracks, showers, and double cascades and their distributions in measured energy and direction.
Where a collaboration has not published a full response, we complete what it has published with the public data of IceCube or KM3NeT/ARCA, moved to the telescope's site and scaled to its size.

\begin{figure*}[t]
  \centering
  \includegraphics[width=\textwidth]{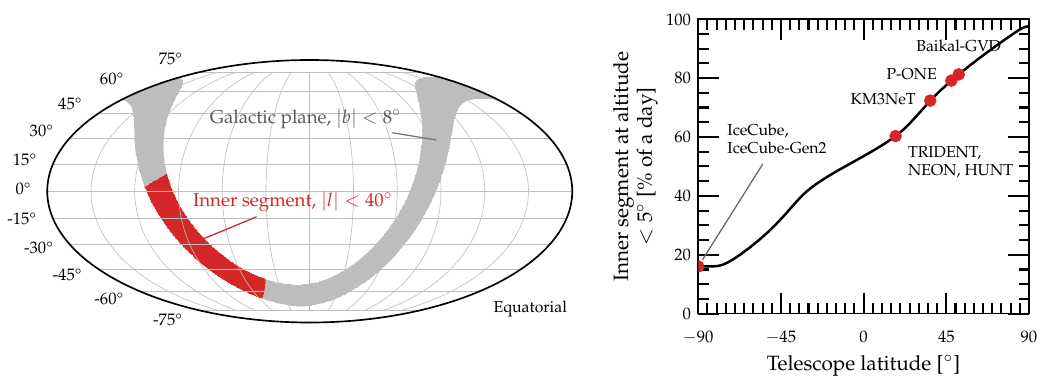}
  \caption{\label{fig:sites}\textbf{\textit{Where the telescopes see the inner segment in tracks.}}
  \textit{Left:} the Galactic plane and the inner segment on the sky.
  \textit{Right:} share of a sidereal day that the inner segment, taken uniform on the sky, spends below $5^\circ$ altitude, where tracks are kept, against the telescope's latitude.
  See the text for details.
  Only the telescopes at the South Pole, IceCube and IceCube-Gen2, see the inner segment in tracks for less than half of the day.}
\end{figure*}

\subsection{Tracks}

For each season and band of declination, IceTracks-DR2 releases the effective area of $\nu_\mu$ and $\bar\nu_\mu$ and, for each neutrino energy, the joint distribution of measured energy and angular error~\cite{Abbasi:2026ehs, IceCube:DR2Data}.
Used as released, these include IceCube's selection and reconstruction and the crossing of Earth, with neutral-current interactions and tau regeneration.
Taus that decay into muons, 17.39\% of them~\cite{ParticleDataGroup:2024cfk}, add tracks from $\nu_\tau$, with the $\nu_\mu$ response read at the neutrino energy that yields the same mean muon energy.
Where a telescope publishes its pointing or energy precision against muon energy, IceCube's is scaled by the ratio of the two at the same muon energy.

\subsection{Showers and double cascades}

Telescopes without published areas for them start from those of IceCube's medium-energy starting events (MESE), released from 1~TeV to 10~PeV for each channel, \ie, flavor, interaction, and event type, and for each band of direction~\cite{IceCube:MESEflavorData}.
Divided by $\ln 10$, as the unstated convention of their energy bins requires, these areas reproduce within 2\% the astrophysical events that IceCube's spectral-break analysis expects at its best fit~\cite{IceCube:2025tgp, IceCube:MESEbreakData}.
Misidentified channels, which few simulated events populate, take their areas averaged over direction.
IceCube's own showers come instead from its Galactic-plane cascade sample (Appendix~\ref{sm:icgp}).

\subsection{Deposited energy}

Each channel deposits its own share of the neutrino energy, $E$, which we express as the energy of an electromagnetic shower that makes as much light.
We draw the share taken by the hadrons, $y$, from the cross sections of \Ref~\cite{Cooper-Sarkar:2011jtt}.
In a $\nu_e$ charged-current interaction, the electron deposits $(1-y)E$ and the hadrons $F(yE)\,yE$, where $F$ is the light yield of hadrons of energy $E_h$ relative to that of an electromagnetic shower~\cite{IceCube:2013llx},
\begin{equation}
  F(E_h) = 1-(1-f_0)\left(\frac{E_h}{E_F}\right)^{-m} ,
  \label{equ:lightyield}
\end{equation}
with $E_F = 0.399$~GeV, $m = 0.130$, and $f_0 = 0.467$.
Neutral-current interactions leave only the hadrons' light; $\nu_\tau$ charged-current ones add the visible part of the tau's decay, with the energy of its $\nu_\tau$ distributed as in \Ref~\cite{Dutta:2000jv}, for the branching fractions of \Ref~\cite{ParticleDataGroup:2024cfk} and a polarization of $-1$.
This model reproduces within 10\% the median measured energy in each channel of IceCube's simulated high-energy starting events (HESE)~\cite{IceCube:2020wum}.

\subsection{Event selections and volumes}

An event selection, the cuts that decide which events enter a sample, keeps a fraction $\epsilon(E_{\rm dep})$ of the showers that deposit $E_{\rm dep}$.
We model it as $\epsilon(E_{\rm dep}) = [1+(E_{50}/E_{\rm dep})^{n}]^{-1}$ (our choice of form), with $E_{50}$ where the selection is half efficient and $n$ setting its steepness.
We fit $E_{50}$, $n$, and the fully efficient area to each published area of $\nu_e$ charged-current showers, band by band in direction, with the absorption in Earth of the PREM model~\cite{Dziewonski:1981xy} where it matters.
For MESE, the selection fitted to its $\nu_e$ showers alone, with our model of deposited energy, predicts the MESE areas of neutral-current and $\nu_\tau$ showers within 10\%.
For a telescope without its own shower areas, each channel's area is MESE's rescaled as
\begin{align}
  \frac{A_{\rm ch}(E,\theta_z)}{A^{\rm MESE}_{\rm ch}(E,\theta_z)} &= \frac{V}{V_{\rm MESE}}
  \label{equ:showers_new} \\
  &\;\times \frac{\int dE_{\rm dep}\,p_{\rm ch}(E_{\rm dep}|E)\,\epsilon(E_{\rm dep})}{\int dE_{\rm dep}\,p_{\rm ch}(E_{\rm dep}|E)\,\epsilon_{\rm MESE}(E_{\rm dep},\theta_z)} ,
  \nonumber
\end{align}
with $\theta_z$ the zenith angle, $p_{\rm ch}$ the distribution of the energy that the channel deposits, $V$ the telescope's instrumented volume, and $V_{\rm MESE}$ the effective volume of MESE's fully efficient selection, 0.27--0.33~km$^3$.
This keeps, for every channel, the Earth crossing of IceCube's simulation.
Outside the energies that a release covers, areas continue with IceCube's energy dependence.

\subsection{Energy precision and pointing}

We smear deposited energies log-normally, in ice with the precision of IceCube's Galactic-plane cascades, from 16\% at 1~TeV to 8--10\% above 10~TeV~\cite{IceCube:2023ame, IceCube:GalacticPlaneData}, and in water with that of KM3NeT/ARCA's showers, 5.5--10.5\%~\cite{KM3NeT:2024paj}.
Point spreads are the distributions of angular errors that IceTracks-DR2 and ARCA release or, elsewhere, King functions~\cite{King:1962wi, Fermi-LAT:2012fsm},
\begin{equation}
  \mathcal{K}(\psi) \propto \left(1-\frac{1}{\gamma_{\rm K}}\right)\left[1+\frac{c^2}{2\gamma_{\rm K}\sigma^2}\right]^{-\gamma_{\rm K}} ,
  \label{equ:king}
\end{equation}
per steradian, with $c = 2\sin(\psi/2)$ for an angle $\psi$ from the neutrino's direction, $\sigma$ the width, and $\gamma_{\rm K}$ setting the tails.
Fermi-LAT normalizes this form in the small-angle approximation~\cite{Fermi-LAT:2012fsm}; written in the chord $c$, it is normalized exactly on the sphere.
We fit $\sigma$ and $\gamma_{\rm K}$ at each energy to published quantiles of the angular error, those of IceCube's Galactic-plane cascades for showers taken from IceCube; only IceCube-Gen2's tracks take both from toise~\cite{toise}.
A telescope that publishes only a median takes ARCA's point spread~\cite{KM3NeT:ARCAirfData} stretched to it, \ie, with every angle multiplied by the ratio of that median to ARCA's at the same energy.

\subsection{Detector sites}

Seen from latitude $\lambda$, a direction at declination $\delta$ moves as $\cos\theta_z = \sin\lambda\sin\delta + \cos\lambda\cos\delta\cos h$, with $h$ the hour angle.
We follow it through a day in 720 steps, reading each response at the local zenith angle, and IceCube's tracks at the declination where IceCube sees that zenith angle, $\theta_z - 90^\circ$.
Tracks count only from $\theta_z > 85^\circ$, where Earth stops atmospheric muons.

Figure~\ref{fig:sites} shows how a telescope's site sets which of the inner Galactic segment's events it sees in tracks.
The segment spans declinations from $-65^\circ$ to $+10^\circ$, so the farther north a telescope lies, the larger the share of each day that the segment spends below $5^\circ$ altitude.
At the South Pole, only its part north of $-5^\circ$ declination lies there, so IceCube sees the inner segment mostly in showers, while the telescopes in the north add its tracks.

\begin{table*}[t]
  \caption{\label{tab:telescopes}\textbf{\textit{The eight neutrino telescopes.}}
  Location, medium, depth of the instrumented volume's center, and the size that sets our shower rates.
  In each event type, ``Own'' is the response that the collaboration has published, ``IceCube'' or ``ARCA'' one taken from that telescope, and ``Own + IceCube'' or ``Own + ARCA'' a combination of the two.}
  \centering
  \begin{ruledtabular}
  \begin{tabular}{lllllll}
    Telescope & Site & Medium, depth & Size & Tracks & Showers & Double cascades \\
    \hline
    IceCube      & South Pole              & Ice, 1.95~km               & 1~km$^3$                       & Own           & Own (plane sample) & --- \\
    KM3NeT/ARCA  & $36^\circ16'$\,N        & Sea, 3.11~km               & 1~km$^3$                       & Own           & Own           & IceCube \\
    Baikal-GVD   & $51^\circ46'$\,N        & Lake, 1.01~km              & 20 clusters\footnote{Our choice: about the 1~km$^3$ that \Ref~\cite{Saurenhaus:2026hzj} takes as Baikal-GVD's full size.} & Own + IceCube & Own           & IceCube \\
    P-ONE        & $47^\circ46'$\,N        & Sea, 2.16~km               & 0.37~km$^3$\footnote{Our choice: P-ONE's 70 strings, each with its share of their 80-m grid, over the 950-m span of their modules~\cite{Twagirayezu:2023cpv}.} & Own + IceCube & ARCA          & IceCube \\
    TRIDENT      & $17.4^\circ$\,N         & Sea, 3.1~km                & 7.5~km$^3$                     & Own           & Own           & Own \\
    NEON\footnote{Our choice of site and depth: in the same sea as TRIDENT~\cite{Zhang:2024slv}, at the deeper of the two depths that NEON considers.} & $17.4^\circ$\,N & Sea, 3.5~km & 6.8~km$^3$ & Own + IceCube & Own + ARCA    & IceCube \\
    HUNT\footnote{Our choice of site and depth: its South China Sea candidate site~\cite{Huang:2023mzt}, with its 2025 clusters on the 3-km seabed~\cite{Huang:2025auf}.} & $17.4^\circ$\,N & Sea, 2.06~km & 34~km$^3$ & Own + IceCube & ARCA          & IceCube \\
    IceCube-Gen2 & South Pole              & Ice, 2.02~km               & IceCube + 4.9~km$^3$           & Own + IceCube & Own + IceCube & Own \\
  \end{tabular}
  \end{ruledtabular}
\end{table*}
\begin{table*}[t]
  \caption{\label{tab:checks}\textbf{\textit{Checks of our responses vs.~published results.}}
  Each row is a published result that we recompute with our response, emulating each collaboration's analysis.
  The third column gives our result over the published one, unless stated otherwise; the fourth, the largest deviation from 1 that we accept, fixed before computing.
  See the text for details.}
  \centering
  \begin{ruledtabular}
  \begin{tabular}{L{2.6cm}L{7.3cm}L{4.4cm}L{2.9cm}}
    Telescope & Published result & Ours / published & Pass mark \\
    \hline
    IceCube & IceTracks-DR2: $5\sigma$ discovery flux in 14~yr at $\sin\delta = 0$--0.9, $E^{-2}$ / $E^{-3}$ & 0.94--1.13 / 0.93--0.97 & 25\% \\
 & Galactic-plane cascades: 68\%~C.L.\ flux intervals in IceCube's regions (Appendix~\ref{sm:icgp}) & Edges within 10 points in 13 of 16; inner Galaxy $\pm32\%$, IceCube's $-29\%/{+32}\%$ & --- \\
 & MESE: astrophysical events of each event type at the best fit of its flavor analysis, and their ratios & 1.14--1.16; ratios within 1.4\% & 10\% on ratios; the offset sets a prior \\
 & HESE: median measured energy of the showers in each channel, 0.1--3~PeV & 0.90--0.96 & 10\% \\
 & PLEnuM~\cite{Schumacher:2025qca}: uncertainties of NGC~1068's flux and spectral index, six networks & 0.89--1.34 (0.88--1.07 with one pointing) & 20\%; one fails\footnotemark[1] \\
KM3NeT/ARCA & $5\sigma$ discovery flux: 4 spectral indices, 3 live times, 9 declinations & 0.76--1.14 (median 0.98) & 25\% \\
Baikal-GVD & Efficiency for $\nu_\tau$ double cascades & 1.05 & 25\% \\
 & Shower areas per flavor above 3.5~PeV & 6 of 14 within 10\% & 10\%; fails\footnotemark[2] \\
P-ONE & None published & --- & --- \\
TRIDENT & Time to discover NGC~1068 at $5\sigma$; 10-yr $5\sigma$ discovery flux, $E^{-2}$ / $E^{-3}$ & 0.98; 1.14 / 1.37 & 25\%; $E^{-3}$ fails\footnotemark[3] \\
NEON & 10-yr $5\sigma$ discovery flux, $\sin\delta = -0.9$ to $-0.6$ & 1.05--1.09 & 25\% \\
HUNT & $5\sigma$ discovery flux of NGC~1068 in 1 and 10~yr, for 12 energy thresholds & 0.93--1.01\footnotemark[4] & 25\% \\
IceCube-Gen2 & Significance of NGC~1068 in 10~yr, relative to IceCube & 1.02 & 25\% \\
 & Showers in the new strings, $\nu_e$ / $\nu_\mu$ / $\nu_\tau$ & 0.89--1.05 / 0.65--1.08 / 1.11--1.45 & 20\%; $\nu_\mu$, $\nu_\tau$ fail\footnotemark[5]

  \end{tabular}
  \footnotetext[1]{The largest network fails because IceTracks-DR2's pointing, read at each site's zenith angles, is wider for up-going tracks than for horizontal ones, while PLEnuM gives all its telescopes one pointing; with one pointing, all pass.}
  \footnotetext[2]{Baikal-GVD derived these areas with a looser selection than ours.}
  \footnotetext[3]{With the response of TRIDENT's 2023 proceedings~\cite{Chang:2023aae}, we reproduce the published $E^{-3}$ flux, so TRIDENT likely computed it with that response.}
  \footnotetext[4]{Measuring muon energies with the precision that HUNT assumes; with IceCube's measured energies, which we keep, 1.06--1.21.}
  \footnotetext[5]{IceCube-Gen2's code counts all the energy of hadrons as visible and drops the hadrons of $\nu_\tau$ charged-current interactions; under its rules, 0.76--1.11 and 0.75--0.99.}
  \end{ruledtabular}
\end{table*}

\subsection{Galactic segments and their bins}

Averaged over a day, a telescope's response depends on the direction of the neutrinos only through its declination, so we compute it at 90 declinations, from $-89^\circ$ to $+89^\circ$.
For the diffuse emission of a segment, we divide the segment into pixels (HEALPix~\cite{Gorski:2004by}, $N_{\rm side} = 64$, in Galactic coordinates) and add their events.
Each pixel takes the response at the nearest of these declinations, weighted by its solid angle.
We bin the events in measured energy and in measured Galactic latitude, $b$, at the segment's longitudes, with edges at $|b| = 0^\circ$, $4^\circ$, $8^\circ$, $12^\circ$, $20^\circ$, and $30^\circ$.
An event from the segment can be reconstructed outside it, especially a shower, whose angular error is large.
The bins from $8^\circ$ to $30^\circ$ collect such events.
The joint fit also counts those reconstructed in the other segments or in the sky away from the plane.
In each bin, the background is the response's background per steradian, read at each declination that the bin spans and multiplied by the bin's solid angle there.

\subsection{The eight telescopes}

Table~\ref{tab:telescopes} lists the eight neutrino telescopes we consider, their sites and sizes, and where the response of each event type comes from.

For KM3NeT/ARCA, we use its released response for 230 strings: the areas of tracks and of $\nu_e$ charged-current showers by direction, the spreads of their measured energies and angular errors, and the backgrounds~\cite{KM3NeT:ARCAirfData}.
Its $\nu_\tau$ charged-current and neutral-current showers and its Glashow resonance~\cite{Glashow:1960zz, IceCube:2021rpz} come from its all-sky areas per channel~\cite{KM3NeT:2024paj}, with the direction dependence of its $\nu_e$ showers; its $\nu_\mu$ showers and double cascades are MESE's, rescaled by \equ{showers_new}.

Baikal-GVD's cascade areas per flavor~\cite{Baikal-GVD:2020irv, Avrorin:2022lyk} predict more events than its analyses, so we scale them down to match its published samples~\cite{GVD:2025lya, Baikal-GVD:2022fis, Baikal-GVD:2021civ}, by 0.63 below 70~TeV and by 0.65--0.97 above.
Its showers have a median pointing of $3^\circ$ and an energy precision of 20\%. 
Its tracks are IceTracks-DR2's, scaled by the projected area of a cluster~\cite{Baikal-GVD:2022fmn} and by a selection in muon energy fitted to its published sample, whose pointing they take~\cite{Safronov:2025dbo}.

P-ONE's tracks are IceTracks-DR2's, scaled by the ratios of its published area and median angular error to IceCube's~\cite{Twagirayezu:2023cpv, Karle:2010xx}.
Its showers take ARCA's selection, precision, and pointing, scaled to its volume.

TRIDENT's tracks and showers take its areas and pointing~\cite{Morton-Blake:2025vpp}, scaled to 1\,211 strings~\cite{TRIDENT:2022hql}, with energy precision~\cite{Wang:2026kka} and double-cascade identification~\cite{Tian:2025ymh}.

NEON's tracks take its area~\cite{Zhang:2024slv}, ARCA's point spread stretched to its median, and IceCube's measured energies; its showers take ARCA's selection, with the pointing and precision of NEON's reconstruction~\cite{Lin:2026zrg}.

HUNT's tracks take its published all-sky area and median angular error~\cite{Huang:2025auf}, with the direction dependence of its clusters' projected area (our assumption); its showers take ARCA's precision, pointing, and selection, with the threshold raised to 10~TeV.

IceCube-Gen2's tracks are IceTracks-DR2's, times the gain over IceCube that the models in its technical design report give~\cite{IceCube-Gen2:TDR, toise, photospline, Whitehorn:2013nh}. 
Its new strings add showers and double cascades, scaled from MESE by the report's selection, whose turn-on is near 67~TeV.

\subsection{Checks}
\label{sm:checks}

Table~\ref{tab:checks} lists our checks: with each response, we redo analyses that its collaboration has published, as the publication describes them, and compare the results.
Five concern discovery fluxes, the fluxes of a point source that the telescope would discover at $5\sigma$.
The few mismatches arise where a published result rests on a simpler model, an earlier response, or a looser event selection than ours. 
In each case, ours follows the latest published data or IceCube's simulation more closely (see the table notes).

\section{IceCube Galactic-plane cascades}
\label{sm:icgp}

IceCube's showers come from the cascade sample of its analyses of the Galactic plane~\cite{IceCube:2023ame, IceCube:2026hqk}: 59\,592 events from May 2011 to May 2021, with measured energies above 500~GeV, selected by deep learning, of which 87\% are atmospheric.
Its release gives one effective area for the three flavors, in five bands of $\sin\delta$, the expected events of each flavor for an isotropic astrophysical flux, the data and the expected backgrounds against declination, and the sample's energy and angular resolutions~\cite{IceCube:GalacticPlaneData, IceCube:2023ame}.

\subsection{Flavors}

Since the release gives one area for the three flavors, we split it by flavor with our model of deposited energies, smeared with the release's energy resolution.
Showers are kept above a measured energy of 500~GeV, with the efficiency $\epsilon(E_{\rm dep})$ of Appendix~\ref{sm:telescopes}.
The selection favors showers, but $\nu_\mu$ charged-current events, which make muon tracks, still account for 13\% of its astrophysical events~\cite{IceCube:2023ame}.
We keep them with the shower efficiency times a factor $k_\mu < 1$.
Fitted to the released events of each flavor, the selection is half efficient at 1.64~TeV, with $n = 0.96$, and $k_\mu = 0.25$. 
Values of $k_\mu$ between about 0.20 and 0.28 are also compatible with the released events, so we give it a Gaussian prior of $\pm16\%$ around 0.25 (Table~\ref{tab:priors}).

\subsection{Backgrounds}

We calibrate the sample's two backgrounds, atmospheric neutrinos and muons, on IceCube's release.
First, down-going atmospheric neutrinos survive the self-veto (Appendix~\ref{sm:stats}) with MESE's probability raised to the power 0.45, which we fit to the shapes of IceCube's simulated atmospheric neutrinos, by flavor against measured energy and against declination.
Second, we scale our atmospheric neutrinos by a smooth function of declination so that all parts together, including the plane's signal at IceCube's best fits, match the released data against declination.
This function lies between about 1.0 and 1.4, except within about $25^\circ$ of the South Pole, where it falls to 0.45.
Third, the release gives the number of atmospheric muons, 3\,840, but not their energies; we take their distribution in measured energy as that of the data minus our neutrinos, below 3~TeV.

\subsection{Pointing}

The data release gives the angular errors against true energy, but we bin events in measured energy.
Since the events in a bin of measured energy come from a range of true energies, we start from the King function of Appendix~\ref{sm:telescopes} at the true energy whose median measured energy is that of the bin.
This King function alone gives point spreads 5--9\% narrower than the release's at 2--20~TeV, so we multiply its width by a smooth function of measured energy.
We fit this function so that, at each true energy, the point spread averaged over the measured energies of its events reproduces the released angles that contain 20\%, 50\%, and 80\% of the events.
From 1~TeV to 1~PeV, our angles match the released ones to 1.6\% rms, and to 3.9\% at most.

\subsection{IceCube's measurement redone}

To test this model, we redo on Asimov data IceCube's fits of the flux and spectral index~\cite{IceCube:2026hqk} in the 16 sky regions where it measures a flux: its segments of the plane and the regions of gamma-ray observatories.
To compare like with like, we estimate the background as IceCube does, from the data averaged in right ascension.
In 13 of the 16 regions, the edges of our 68\% flux intervals at 5~TeV, relative to the best fit, lie within 10 percentage points of IceCube's (Table~\ref{tab:checks}).
Alone, the sample measures no flavor, since all its events look like showers.
    
\section{Backgrounds, systematics, statistics}
\label{sm:stats}

\subsection{Atmospheric neutrinos}

We use MCEq~\cite{Fedynitch:2015zma}, as IceCube did in its observation of the Galactic plane~\cite{IceCube:2023ame}, to compute the conventional and prompt fluxes of $\nu_e$, $\nu_\mu$, $\bar\nu_e$, and $\bar\nu_\mu$, from cosmic rays as in H4a~\cite{Gaisser:2011klf} and hadronic interactions as in Sibyll-2.3d~\cite{Riehn:2019jet}.
We pass each neutrino species through its own response.
Following IceCube~\cite{IceCube:2025ewu}, their uncertainties come from pion and kaon production in air showers~\cite{Barr:2006it} and from the spectral index and composition of cosmic rays, between H4a and GST4~\cite{Gaisser:2013bla}.

Down-going atmospheric neutrinos that interact inside IceCube often arrive with muons from their air shower, which make the selection reject them~\cite{Schonert:2008is}.
We compute the fraction that survives this self-veto with nuVeto~\cite{Arguelles:2018awr}, using MESE's probability that a muon of a given energy triggers the veto~\cite{Basu:2025thesis}.
Appendix~\ref{sm:icgp} adapts it to IceCube's plane sample.
We apply no self-veto to the other telescopes: KM3NeT/ARCA and Baikal-GVD use the backgrounds that their collaborations publish.
The rest keep showers only from below or near the horizon.

\subsection{Atmospheric muons}

Atmospheric muons enter only the shower samples, since tracks are kept only from directions where Earth stops the muons (Appendix~\ref{sm:telescopes}).
IceCube's come from its data (Appendix~\ref{sm:icgp}).
P-ONE, TRIDENT, NEON, and HUNT take KM3NeT/ARCA's published muons~\cite{KM3NeT:2024paj}, scaled by their volume and by the intensity of muons at their depth~\cite{Bugaev:1998bi}; IceCube-Gen2's new strings take MESE's, scaled in the same way and by their selection.

\subsection{Systematic uncertainties}

Table~\ref{tab:priors} lists the model's parameters, their true values, and their priors.
Where a collaboration states an uncertainty, we take it as the prior; otherwise, we set the prior from our checks (Table~\ref{tab:checks}) or assume it, as the table's last column records.
IceCube's released pointing for showers leaves out the uncertainties in the detector's properties, chiefly the optical properties of the ice and the efficiency of its light sensors.
IceCube estimates that these widen the angular errors by about 5\% at 1~TeV and 30\% at 1~PeV~\cite{IceCube:2023ame}.
We widen it by a fraction $f(E)$ that grows linearly in $\log_{10}E$ from 5\% to 30\% between these energies and stays constant outside, as
\begin{equation}
  \psi = \psi_{\rm rel}\left[1 + f(E)\,(1+t)\right] ,
  \label{equ:dnn}
\end{equation}
with $\psi_{\rm rel}$ the released angular error and $t$ a nuisance parameter with prior $0 \pm 1$; $t = -1$ recovers $\psi_{\rm rel}$.

\subsection{Statistics}

The fit minimizes the function
\begin{equation}
  \chi^2 = 2\sum_i\left[\mu_i - n_i + n_i\ln\frac{n_i}{\mu_i}\right] + \sum_k \frac{(\eta_k - \bar\eta_k)^2}{\sigma_k^2} ,
  \label{equ:chi2}
\end{equation}
where $i$ runs over the bins of every telescope and event type (Appendices~\ref{sm:telescopes} and~\ref{sm:regions}), $n_i$ are the Asimov data, \ie, the events expected from the truth, $\mu_i$ are the events expected for given values of the model parameters, and $\eta_k$ are the nuisance parameters with Gaussian priors of mean $\bar\eta_k$ and width $\sigma_k$; parameters with uniform priors are only bounded.

For the parameters that we measure, such as the flavor fractions at Earth of a segment, we profile $\chi^2$: at each of their values, we minimize $\chi^2$ over all other parameters.
By Wilks' theorem, the 68\% and 95\%~C.L.\ regions of the two independent fractions at Earth, \eg, $\fE{e}$ and $\fE{\mu}$, lie where $\Delta\chi^2 < 2.30$ and 6.18, and the one-dimensional intervals of $\fS{e}$ where $\Delta\chi^2 < 1$ and 3.84.
We keep each flavor fraction between 0 and 1.
The mixing parameters are fixed at the best fit of NuFIT~6.1~\cite{Esteban:2024eli, NuFIT61}, except in the intervals of $\fS{e}$ (\figu{fes}) and the fits of neutrino decay, where they vary with Gaussian priors centered there, of the widths projected for 2040~\cite{Bustamante:2026fxa}.

\begin{table*}[t]
  \caption{\label{tab:priors}\textbf{\textit{Model parameters.}}
  True values are those of the Asimov data; priors are Gaussian unless stated.
  The plane's values are for S1, S2, and S3; S3's index is IceCube's lower limit.
  Normalizations are in $10^{-11}$~TeV~cm$^{-2}$~s$^{-1}$~sr$^{-1}$.
  A parameter marked ``each telescope its own'' has one copy per telescope; any other is shared by the telescopes in its row.
  The mixing parameters stay at their true values except in the intervals of $\fS{e}$ and the fits of neutrino decay.}
  \centering
  \begin{ruledtabular}
  \begin{tabular}{L{6.2cm}L{3.6cm}L{2.2cm}L{1.9cm}L{3.3cm}}
    Parameter & Applies to & True value & Prior & Source \\
    \hline
    \textit{Galactic plane (each segment its own)} & & & & \\
Normalization, $E^2\Phi$ per flavor, 5~TeV & All & 19.1, 5.1, 3.5 & Free & \cite{IceCube:2026hqk} \\
Spectral index & All & 2.83, 2.93, 2.87 & Free & \cite{IceCube:2026hqk} \\
Flavor composition & All & Pion decay & Free & --- \\
\textit{Atmospheric $\nu$ (all telescopes)} & & & & \\
Conventional atmos.\ $\nu$, normalization & All & 1 & $\pm25\%$ & \cite{IceCube:2025ewu} \\
Prompt atmos.\ $\nu$, normalization & All & 1 & Free, 0--4 & \cite{IceCube:2025ewu} \\
Cosmic-ray spectral index, shift & All & 0 & $\pm0.055$ & \cite{IceCube:2025ewu} \\
Cosmic-ray composition, H4a (0)--GST4 (1) & All & 0 (H4a) & $\pm1$ & \cite{IceCube:2025ewu} \\
$\pi$ production in air showers, Barr H & All & 0 & $\pm0.15$ & \cite{IceCube:2025ewu, Barr:2006it} \\
$K$ production in air showers, Barr W & All & 0 & $\pm0.4$ & \cite{IceCube:2025ewu, Barr:2006it} \\
$K$ production in air showers, Barr Y & All & 0 & $\pm0.3$ & \cite{IceCube:2025ewu, Barr:2006it} \\
$K$ production in air showers, Barr Z & All & 0 & $\pm0.12$ & \cite{IceCube:2025ewu, Barr:2006it} \\
\textit{Extragalactic flux (all telescopes)} & & & & \\
Normalization, $E^2\Phi$ per flavor, 100~TeV & All & 1.77 & $\pm10.5\%$ & \cite{IceCube:2025ewu} \\
Index below break & All & 1.31 & $\pm0.905$ & \cite{IceCube:2025ewu} \\
Index above break & All & 2.735 & $\pm0.071$ & \cite{IceCube:2025ewu} \\
$\log_{10}$(break energy/GeV) & All & 4.39 & $\pm0.1$ & \cite{IceCube:2025ewu} \\
Flavor composition & All & Pion decay & Free & --- \\
\textit{Shared by several telescopes} & & & & \\
IceTracks-DR2's normalization & IceCube, P-ONE, Gen2 & 1 & $\pm11\%$ & \cite{IceCube:2016tpw} \\
IceTracks-DR2's energy scale & IceCube, P-ONE, NEON, HUNT, Gen2 & 1 & $\pm5\%$ & \cite{IceCube:2025ewu} \\
Deposited-energy model, energy scale & All & 1 & $\pm10\%$ & Our checks (Table~\ref{tab:checks}) \\
NC deposited energy, relative to $\nu_e$ CC & All & 1 & $\pm5\%$ & Our checks (Table~\ref{tab:checks}) \\
$\nu_\tau$ CC deposited energy, relative to $\nu_e$ CC & All & 1 & $\pm5\%$ & Our checks (Table~\ref{tab:checks}) \\
Misidentified events & All & 1 & $\pm50\%$ & Assumed \\
\textit{Each telescope} & & & & \\
IceCube's plane sample, normalization & IceCube & 1 & $\pm9.4\%$ & Our choice, after~\cite{IceCube:2016tpw} \\
IceCube's plane sample, $\nu_\mu$ kept as showers & IceCube & 1 & $\pm16\%$ & Our checks, after~\cite{IceCube:2023ame} \\
IceCube's shower energy scale & IceCube, Gen2 & 1 & 0.9--1.1, uniform & \cite{IceCube:2025ewu} \\
IceCube's shower pointing, $t$ in \equ{dnn} & IceCube, Gen2 & 0 & $\pm1$ & \cite{IceCube:2023ame} \\
IceCube's track pointing & IceCube, Gen2 & 1 & $\pm10\%$ & Assumed \\
IceCube-Gen2's new strings, efficiency & Gen2 & 1 & $\pm11\%$ & After~\cite{IceCube:2016tpw, IceCube-Gen2:TDR} \\
KM3NeT's efficiency & ARCA & 1 & $\pm10\%$ & \cite{KM3NeT:2024uhg, KM3Net:2016zxf} \\
KM3NeT's energy scale & ARCA & 1 & $\pm10\%$ & Assumed, after~\cite{KM3NeT:2024paj} \\
KM3NeT's pointing & ARCA & 1 & $\pm10\%$ & Assumed, after~\cite{KM3NeT:2024paj} \\
Efficiency, each telescope its own & P-ONE, TRIDENT, NEON, HUNT & 1 & $\pm10\%$ & ARCA's~\cite{KM3NeT:2024uhg} \\
Energy scale, each telescope its own & P-ONE, TRIDENT, NEON, HUNT & 1 & $\pm10\%$ & ARCA's~\cite{KM3NeT:2024paj} \\
Pointing, each telescope its own & P-ONE, TRIDENT, NEON, HUNT & 1 & $\pm10\%$ & ARCA's~\cite{KM3NeT:2024paj} \\
Baikal-GVD's shower efficiency & GVD & 1 & $\pm22.4\%$ & \cite{GVD:2025lya, Baikal-GVD:2022fis} \\
Baikal-GVD's shower energy scale & GVD & 0 & $\pm0.05$, $\log_{10}E$ & \cite{GVD:2025lya} \\
Baikal-GVD's shower pointing & GVD & 1 & $\pm10\%$ & Assumed, after~\cite{GVD:2025lya} \\
Baikal-GVD's track efficiency & GVD & 1 & $\pm35\%$ & \cite{GVD:2026cep} \\
Baikal-GVD's track energy scale & GVD & 1 & $\pm5\%$ & Assumed, after~\cite{IceCube:2025ewu} \\
Baikal-GVD's track pointing & GVD & 1 & $\pm10\%$ & Assumed, after~\cite{GVD:2026cep} \\
Atmos.\ $\mu$ in showers, each telescope its own & All & 1 & $\pm52.4\%$ & \cite{IceCube:2025ewu} \\
\textit{Mixing parameters} & & & & \\
$\sin^2\theta_{12}$, $\sin^2\theta_{23}$, $\sin^2\theta_{13}$, $\delta_{\rm CP}$ [$^\circ$] & All & 0.3088, 0.470, 0.02248, 212 & $\pm$0.0016, 0.0042, 0.00059, 6.7 & \cite{Esteban:2024eli, NuFIT61, Bustamante:2026fxa}

  \end{tabular}
  \end{ruledtabular}
\end{table*}

\section{Galactic segments, extragalactic flux}
\label{sm:regions}

\subsection{Six Galactic segments}

IceCube measured the Galactic plane, at $|b| < 8^\circ$, in two, three, and six segments of Galactic longitude, fitting a power law to each~\cite{IceCube:2026hqk}.
Our main results use its three: the inner segment, S1, at $|l| < 40^\circ$, and the outer ones, S2 at $40^\circ < l < 180^\circ$ and S3 at $-180^\circ < l < -40^\circ$. 
Below, we present results for six segments.

Of IceCube's six segments, the inner one lies at $|l| < 30^\circ$.
For outer segments S5 and S6, where IceCube gives only upper limits on the flux, we take the true flux to be that of CRINGE~\cite{Schwefer:2022zly} (see below), averaged over the segment, instead of IceCube's best fit.
Fitting the six segments instead of the three, the inner one is measured as precisely: its $\fS{e}$ to within 26\% with all eight telescopes and 89\% with the existing ones, against 27\% and 88\% for S1 (68\%~C.L.).
The five fainter segments are measured more loosely, with 68\% intervals of $\fS{e}$ from 0.16--0.52 in S6 to 0--0.90 in S5.
The extragalactic $\fS{e}$, measured mostly away from the plane, is unchanged.

\subsection{CRINGE fluxes as the truth}

CRINGE models the Galactic-plane neutrinos from cosmic-ray collisions with interstellar gas, with the cosmic-ray spectra fitted to measurements, plus those of unresolved sources~\cite{Schwefer:2022zly}.
In S1, S2, and S3, its flux at 5~TeV is 0.58, 0.72, and 0.94 times IceCube's~\cite{IceCube:2026hqk}, with a harder spectrum, of index about 2.56 against IceCube's 2.83--2.93.
About half of its flux at 5~TeV comes from the unresolved sources.
With CRINGE fluxes as the truth, the inner segment's $\fS{e}$ is 0.235--0.435 with all eight telescopes, against 0.245--0.425 with IceCube's (68\%~C.L.).

\subsection{Uniformly bright segments}

Inside each segment, we take the neutrino signal as uniform on the sky, as IceCube does in its fits~\cite{IceCube:2026hqk}.
In contrast, models of cosmic-ray collisions with gas concentrate it towards the plane's midline.
We fit the inner one of six segments alone, with its signal following the CRINGE sky map or the $\pi^0$ map of IceCube's 2023 analysis~\cite{IceCube:2023ame, IceCube:GalacticPlaneData, Fermi-LAT:2012edv}, scaled to IceCube's mean flux in the segment.
About 80\% of the signal events measured in it then lie at $|b| < 4^\circ$, against 52\% with the uniform sky.
Since the atmospheric background does not follow the plane, the signal-to-background ratio there grows by about 60\%.
With 10 years of all eight telescopes, $\Delta\chi^2$ against neutron decay grows by about half.
Yet the 68\%~C.L. region of the segment's composition at Earth shrinks by at most 0.05 in each flavor fraction.

\subsection{The sky away from the plane}

We bin the sky at $|b| > 30^\circ$ in ten bands equal in $\sin\delta$, since the responses depend on declination (Appendix~\ref{sm:telescopes}).

\section{Composition varying along the plane}
\label{sm:mixed}

We test whether the telescopes would notice the inner segment making its neutrinos differently from the other two, \ie, with a different flavor composition.
In our test case, the sources in S1 make neutrinos by muon damping, $(0, 1, 0)_{\rm S}$, which becomes $(0.22, 0.41, 0.38)_\oplus$ at Earth, while those in S2 and S3 make them by pion decay.
Each segment has the flux and spectral index that IceCube measured.
We compare two hypotheses with the fit of these data in which every segment's composition is free.

First, fixing the composition of S1 at that of pion decay, with those of S2 and S3 free, raises $\chi^2$ by 12.7 with all eight telescopes.
This rise rejects pion decay in S1 at $3.1\sigma$ (two degrees of freedom, one per independent flavor fraction fixed).
With the existing telescopes, $\chi^2$ rises by 1.3, or $0.6\sigma$.
Second, letting the three segments share one composition, free, raises $\chi^2$ by only 1.8 with all eight telescopes, or $0.3\sigma$ (four degrees of freedom).
The shared composition is close to that of S1, the brightest segment, and fits the fainter, loosely measured S2 and S3 nearly as well as pion decay does.
Thus, all eight telescopes could tell that S1 does not make its neutrinos by pion decay, but not that the composition varies along the plane.

\section{Cooling and decay: further details}
\label{sm:coolingdecay}

\begin{figure}[t]
  \centering
  \includegraphics[width=\columnwidth]{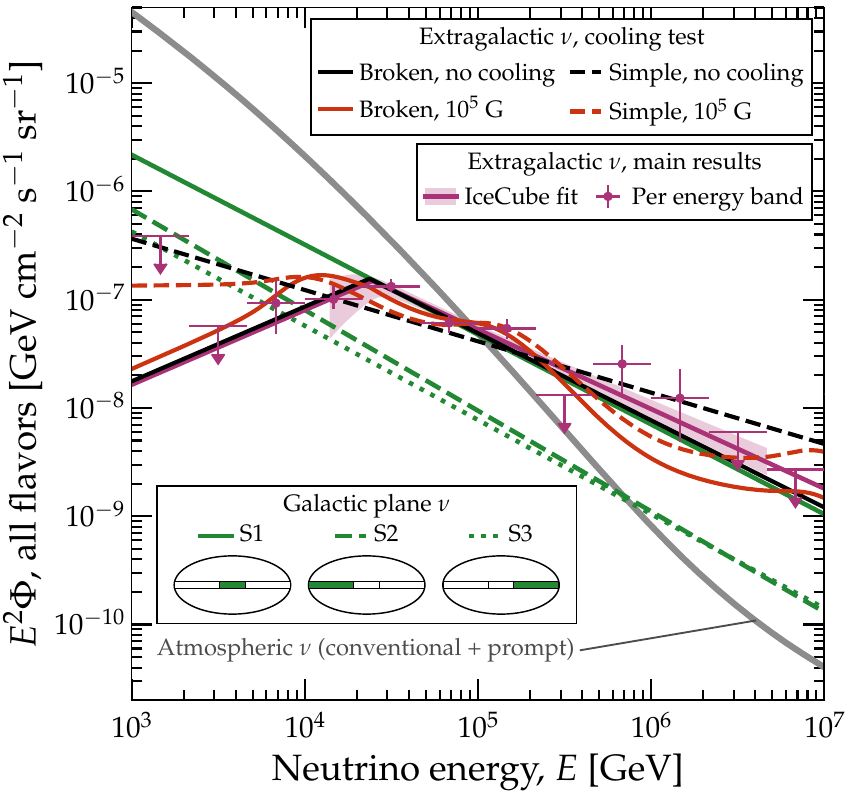}
  \caption{\label{fig:spectra}\textbf{\textit{All-flavor neutrino intensities in the fits.}}
  For the Galactic plane, IceCube's measured flux and spectral index in each segment~\cite{IceCube:2026hqk}, extended as power laws.
  For the extragalactic flux, IceCube's fit, with our estimate of its 68\%~C.L.\ band where it is measured, and IceCube's flux in independent energy bands (68\%~C.L.; arrows for upper limits)~\cite{IceCube:2025ewu}.
  For the cooling test, the sources emit a broken or a simple power law, fitted to those bands.
  For the atmospheric neutrinos, the flux averaged over the sky, before any veto.
  See the text for details.
  Per steradian, every segment outshines the extragalactic flux of our main results below about 10~TeV, and the inner segment up to about 30~TeV.}
\end{figure}

\subsection{Cooling}

We follow pions, kaons, and muons in the extragalactic sources, in a steady state between injection, synchrotron losses, and decay, so that cooled particles pile up below their critical energy, where they decay~\cite{Kachelriess:2007tr, Hummer:2010ai}.
The critical energy grows as $(m^5/\tau)^{1/2}$, as in \equ{ecmu}, so that those of pions and kaons are 18 and 630 times that of muons.
We take 0.1 kaons per pion at the same energy and keep only the decay $K^+ \to \mu^+ \nu_\mu$, which yields $\nu_\mu$ above the pions' critical energy; both choices are ours.

Cooling acts on the spectrum that the sources emit, which we take as a simple or a broken power law.
At each field, we fit the emitted spectrum so that, once cooled, it matches IceCube's flux in 13 independent energy bands from 1~TeV to 100~PeV~\cite{IceCube:2025ewu}.
For the broken power law, we fit only the normalization and the index above the break.
Since the field does not set where the emitted spectrum breaks, we keep the index below the break at 1.31, as in IceCube's fit~\cite{IceCube:2025ewu}, and the break at 24~TeV, where our fit without cooling places it.

Refitting the break at each field instead moves it to absorb part of the bend that cooling makes.
The significance above $3 \times 10^4$~G then falls by up to $2.5\sigma$, while the fields at which all eight telescopes reach $3\sigma$ and $5\sigma$ stay the same.
We stop at $3 \times 10^5$~G, where $E_c \approx 6$~TeV.
At stronger fields, pions also cool among the detected energies and bend the spectrum again (see, \eg, \Ref~\cite{Bustamante:2020bxp}), which neither null hypothesis can follow.

Figure~\ref{fig:spectra} shows the intensities of the plane's segments, of the atmospheric flux, and of the extragalactic flux, both IceCube's fit with its flux per energy band and the true fluxes of the cooling test, without cooling and at $10^5$~G.
We estimate the band of IceCube's fit from the uncertainties of its four parameters~\cite{IceCube:2025ewu}, taken as independent.
Without cooling, the simple power law exceeds the flux that IceCube measured below about 5~TeV.
The pile-ups of muons, pions, and kaons below their critical energies appear as bumps in the cooled spectra.

\subsection{Fitting the field}

Since the field is undefined under the null hypothesis, we convert $\Delta\chi^2$ to a significance with pseudo-experiments: 1\,000 per null hypothesis and set of telescopes, drawn from the null's expected event numbers.
Their $\Delta\chi^2$ exceeds a given value no more often than $\chi^2$ with one degree of freedom, so we use its $p$-values.
Above $\Delta\chi^2 \approx 7$, which too few of the 1\,000 pseudo-experiments reach to test this conversion, we extrapolate it.

\subsection{Decay}

The density of extragalactic sources grows as $(1+z)^{3.4}$ up to $z = 1$ and as $(1+z)^{-0.3}$ from there up to $z = 4$~\cite{Hopkins:2006bw, Yuksel:2008cu, Kistler:2009mv}, in a flat $\Lambda$CDM cosmology with $H_0 = 67.4$~km~s$^{-1}$~Mpc$^{-1}$ and $\Omega_m = 0.315$~\cite{Planck:2018vyg}; each source emits IceCube's measured spectrum.
The plane's sources fill an exponential disc with a scale length of 3~kpc and a height of 0.1~kpc (our choice); we average the survival of their neutrinos over each segment's directions and along each line of sight.
Where decay is detected, it is detected through flavor: up to 10~s~eV$^{-1}$, $\Delta\chi^2$ from the full effect is within 2\% of that from the change in flavor alone, with IceCube's measured spectrum kept (10 years of IceCube).
When $\nu_3$ alone decays, the extragalactic flux reaches Earth with $(0.47, 0.29, 0.24)_\oplus$, closer to that of pion decay than the $(0.68, 0.11, 0.21)_\oplus$ of $\nu_1$.

\subsection{Visible decay}

If $\nu_2$ and $\nu_3$ decay into $\nu_1$ and a scalar, the daughters are detected at a fraction of their parents' energy set by the neutrino masses~\cite{Abdullahi:2020rge}.
Taking neutrinos to be Majorana particles, we profile the lightest mass between 0 and 0.08~eV, penalized by the cosmological bound on the neutrino-mass sum~\cite{Planck:2018vyg}.
For complete decay, at $\tau/m = 1$~s~eV$^{-1}$, $\Delta\chi^2$ vs.~no decay is 669, vs.~666 for invisible decay (all eight telescopes, 10 years of IceCube).

\section{Timeline of the measurement}
\label{sm:when}

\begin{figure}[t]
  \centering
  \includegraphics[width=\columnwidth]{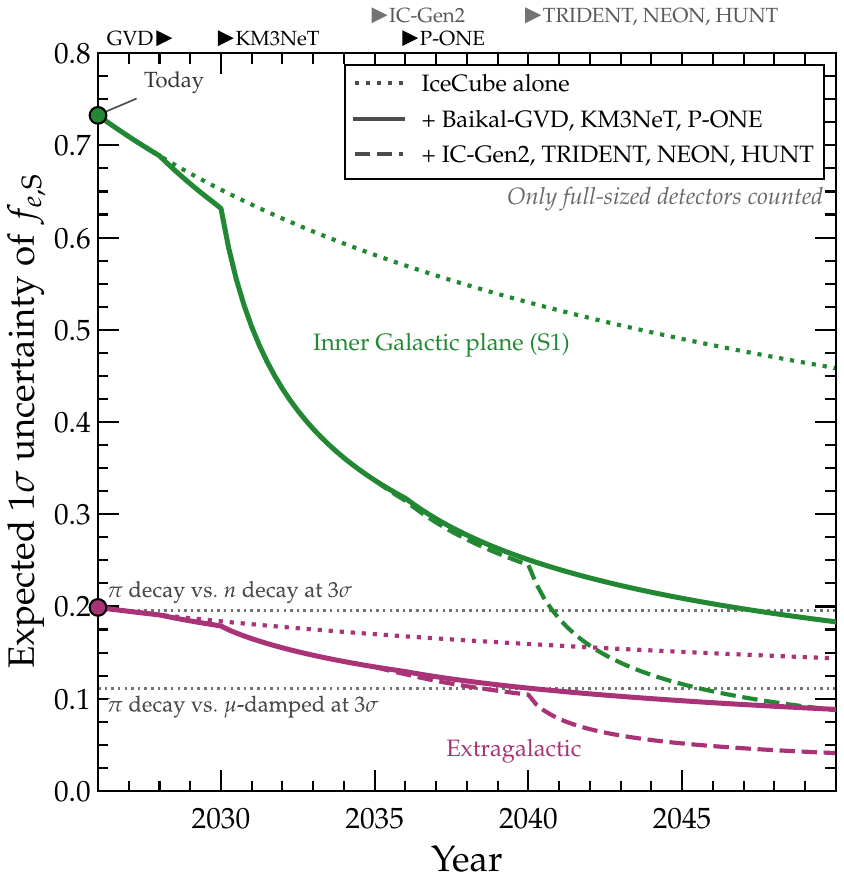}
  \caption{\label{fig:timeline}\textbf{\textit{Expected uncertainty of the $\nu_e$ fraction at the sources, from today to 2050.}}
  Pion decay is the truth for the inner segment and for the extragalactic flux; each telescope counts from its published start at full size~\cite{Saurenhaus:2026hzj, Bustamante:2026fxa}.
  See the text for details.}
\end{figure}

Figure~\ref{fig:timeline} shows the expected $1\sigma$ uncertainty of $\fS{e}$ in each year, for the inner segment and for the extragalactic flux.
We compute it from the curvature of the joint fit's $\chi^2$ at the truth, for sources without $\nu_\tau$ and with fixed mixing parameters, weighting each telescope by its years of data.
IceCube has 15 years of data today, as in our main results.
We count only full-sized telescopes, from published start years.
Baikal-GVD, KM3NeT, and P-ONE count from 2028, 2030, and 2036~\cite{Saurenhaus:2026hzj}.
IceCube-Gen2, which includes IceCube, counts from 2035, and TRIDENT, NEON, and HUNT from 2040~\cite{Bustamante:2026fxa}.

Muon damping, $\fS{e} = 0$, lies half as far from pion decay as neutron decay, $\fS{e} = 1$, so excluding it at the same significance takes twice the precision.
Neutron decay, which yields only $\bar\nu_e$, is rejected by the joint fit with 0.88 times the significance it would have with equal $\nu$ and $\bar\nu$ fluxes (0.90 with all eight telescopes).
Therefore, excluding it needs a 10--12\% smaller uncertainty.
IceCube alone does not reach $3\sigma$ against either alternative in the inner segment by 2050, since it sees the inner Galaxy mainly in showers.
The uncertainty falls steeply once KM3NeT starts, since it sees the inner Galaxy mainly in tracks.
IceCube and KM3NeT alone, on the same schedules, would reach $2\sigma$ against neutron decay in 2038, but not $3\sigma$ by 2050.
With Baikal-GVD and P-ONE, the four reach $3\sigma$ in 2047; all eight telescopes reach it by 2041.

\end{document}